\documentclass[12pt]{article}
 \pdfoutput=1
\usepackage{amssymb}
\usepackage[utf8]{inputenc}
\usepackage{ulem}
\usepackage{xcolor}
\usepackage[pdftex]{graphicx}
\usepackage{epsfig}
\usepackage{cite}
\usepackage{hyperref}
\newcommand{\AmS}{{\protect\the\textfont
  A\kern-.1667em\lower.5ex\hbox{M}\kern-.125emS}}
\newcommand{\ba}{\begin{array}}
\newcommand{\ea}{\end{array}}

\def\beq{\begin{equation}}
\def\eeq{\end{equation}}

\def\beq{\begin{equation}}   
\def\eeq{\end{equation}}

\def\bea{\begin{eqnarray}}
\def\eea{\end{eqnarray}}

\begin{document}

\begin{titlepage}

\begin{flushright}
 IFIC/21-42, FTUV-21-10-12
\end{flushright}

\begin{center}
\vspace{2.7cm}
{\Large{\bf
Searching for hidden matter with long-range angular correlations at ${\bf e^+e^-}$ colliders}}
\end{center}

\vspace{1cm}

\begin{center}

{\bf Redamy P\'erez-Ramos$^{\rm a,\dagger}$ , Miguel-Angel Sanchis-Lozano$^{\rm b,\ddagger}$, and 
Edward K. Sarkisyan-Grinbaum$^{\rm c,d,\star}$}
\vspace{1.5cm}\\

{\it $^a$ DRII-IPSA, Bis, 63 Boulevard de Brandebourg, 94200 Ivry-sur-Seine France}\\
{\it LPTHE, Sorbonne Universit\'e, UPMC Univ Paris 06, UMR 7589,  F-75005, Paris, France} \\

\it $^{\rm b}$ Instituto de F\'{\i}sica
Corpuscular (IFIC) and Departamento de F\'{\i}sica Te\'orica \\
{\it Centro Mixto Universitat de Val\`encia-CSIC, Dr. Moliner 50, E-46100 Burjassot, Spain}

{\it $^{\rm c}$  Experimental Physics Department, CERN, 1211 Geneva 23, Switzerland}

{\it $^{\rm d}$  Department of Physics, The University of Texas at Arlington, Arlington, TX 76019, USA}

\end{center}

\vspace{0.5cm}

\begin{abstract}
The analysis of azimuthal correlations in multiparticle production can be useful to uncover the existence of new 
physics beyond the Standard Model, e.g., Hidden Valley, in $e^+e^-$ 
annihilation at high energies. 
In this paper, based on previous theoretical studies and using PYTHIA8 
event generator, 
it is found that  both azimuthal and rapidity long-range correlations 
are enhanced due to the presence of a new stage of matter on top of the QCD partonic cascade. Ridge structures, similar to those observed at the LHC, 
show up providing a possible signature of new physics 
at future $e^+e^-$ colliders.

\end{abstract}

\begin{center}


\end{center}
\vskip 2.2cm

\begin{small}
\noindent 
$^\dagger$E-mail address: redamy.perez-ramos@ipsa.fr\\
$^\ddagger$E-mail address: Miguel.Angel.Sanchis@ific.uv.es \\ 
$^\star$E-mail address: Edward.Sarkisyan-Grinbaum@cern.ch
\end{small}

\end{titlepage}

In a series of previous papers,
particle correlations were used to extract information about intra-jet multiparticle dynamics 
\cite{Perez-Ramos:2011jic}, and for the search of new physics beyond the 
Standard Model (SM) in hadronic high-energy collisions \cite{Sanchis-Lozano:2008zjj,Sanchis-Lozano:2018wpz}.
Indeed, the analysis of long-range particle correlations can provide useful information about
the early stage of matter 
and shed light on the possible presence of new physics on top of the QCD parton shower. On the other hand, $e^+e^-$ collisions should provide to this end a much cleaner environment than hadron colliders, as well as a definite (though adjustable)
center-of-mass energy. 

 The analyses of 2-particle 
angular correlations in high-multiplicity proton-proton, 
proton-nucleus 
and heavy-ion collisions, 
 have revealed a ridge structure of final state hadrons emitted  
almost collinearly 
 within a broad rapidity difference~\cite{ATLAS:2015hzw,CMS:2010ifv,PHOBOS:2008yxa}. 
Different theoretical explanations~\cite{Shuryak:2007fu,Dumitru:2008wn,Bozek:2012gr} 
have been put forward to explain this initially unexpected phenomenon, almost of all them requiring the existence of some unconventional state of matter at the beginning of the collision. Analogies can be found in the Cosmic Microwave Background \cite{Sanchis-Lozano:2020iwv} where the inflationary epoch plays the role of a hidden sector.

This paper mainly focuses on QCD-like hidden sectors within Hidden Valley (HV) models, where
new types of particles, e.g. v-quarks, undergo new interactions mediated by HV photons ($\gamma_v$) 
or HV gluons ($g_v$) 
\cite{Carloni:2010tw}. 
Production of HV matter will
ultimately enhance and enlarge azimuthal correlations of final-state particles as, e.g., 
in heavy-ion collisions where 
the quark gluon plasma is assumed to be produced~\cite{ShuryakBook}.

The simplest way of coupling the SM and the Hidden Sector (HS) 
occurs via a heavy $Z_v$ of mass $\gtrsim 1$ TeV.
Another possibility involves mirror partners of the SM charged quarks and leptons (collectively denoted as $F_v$) under 
both SM and HS, thereby with the capacity of connecting both sectors. 
Hence $F_v$ can be pair-produced via the annihilation process $e^+e^- \to \gamma/Z^0 \to F_v \overline{F_v}$ above energy threshold. 
The current paper only concerns the lightest and heaviest
cases, namely $D_v$ ($\simeq 100$ GeV) and $T_v$ (several 100 GeV).  
Intermediate masses lead to intermediate results and will not be explicitly reported.

Three kinds of partonic showers, ultimately yielding final-state SM particles, 
are distinguished here:
\begin{itemize}
    \item[(i)] $e^+e^- \to q \bar{q}\to$ hadrons, where $q$ collectively denotes all quark 
flavors below the top ($t$) mass. 
 Using PYTHIA8 event
generator~\cite{Sjostrand:2014zea,Sjostrand:2006za}, 
 all 
flavors are generated 
according to the respective production cross sections.
    \item[(ii)] $e^+e^- \to t \bar{t}\to$ hadrons.
 Top quark possesses special
interest 
 because of its very large mass and the fact that it does not form bound 
states but promptly decays into lighter particles, mimicking the extra cascade due 
to an extra hidden sector. 
    \item[(iii)] $e^+e^- \to F_v \overline{F_v}\to$ hadrons~\cite{Carloni:2010tw}, i.e.
HV production, the main goal of this study. Let us stress that decay modes like $g_v\to g_v\bar{g}_v$, $g_v\to q_v\bar{q}_v$ and $\gamma_v\to q_v\bar{q}_v$ are similar to SM processes along the partonic cascade before hadronization. 
\end{itemize}


The current 
study rests on two 
bases: (a) a previous theoretical study of 2- and 3-particle correlations developed by two 
authors \cite{Sanchis-Lozano:2018wpz,Sanchis-Lozano:2017aur,Sanchis-Lozano:2016qda}, where the possible HV 
contribution  
beyond the SM 
 is incorporated on top of the 
 parton shower, 
and (b) 
 a Monte Carlo study using 
PYTHIA8~\cite{Sjostrand:2014zea,Sjostrand:2006za} with 
different HV sectors  
incorporated.


\begin{figure}[!htbp]
\begin{center}
\epsfig{file=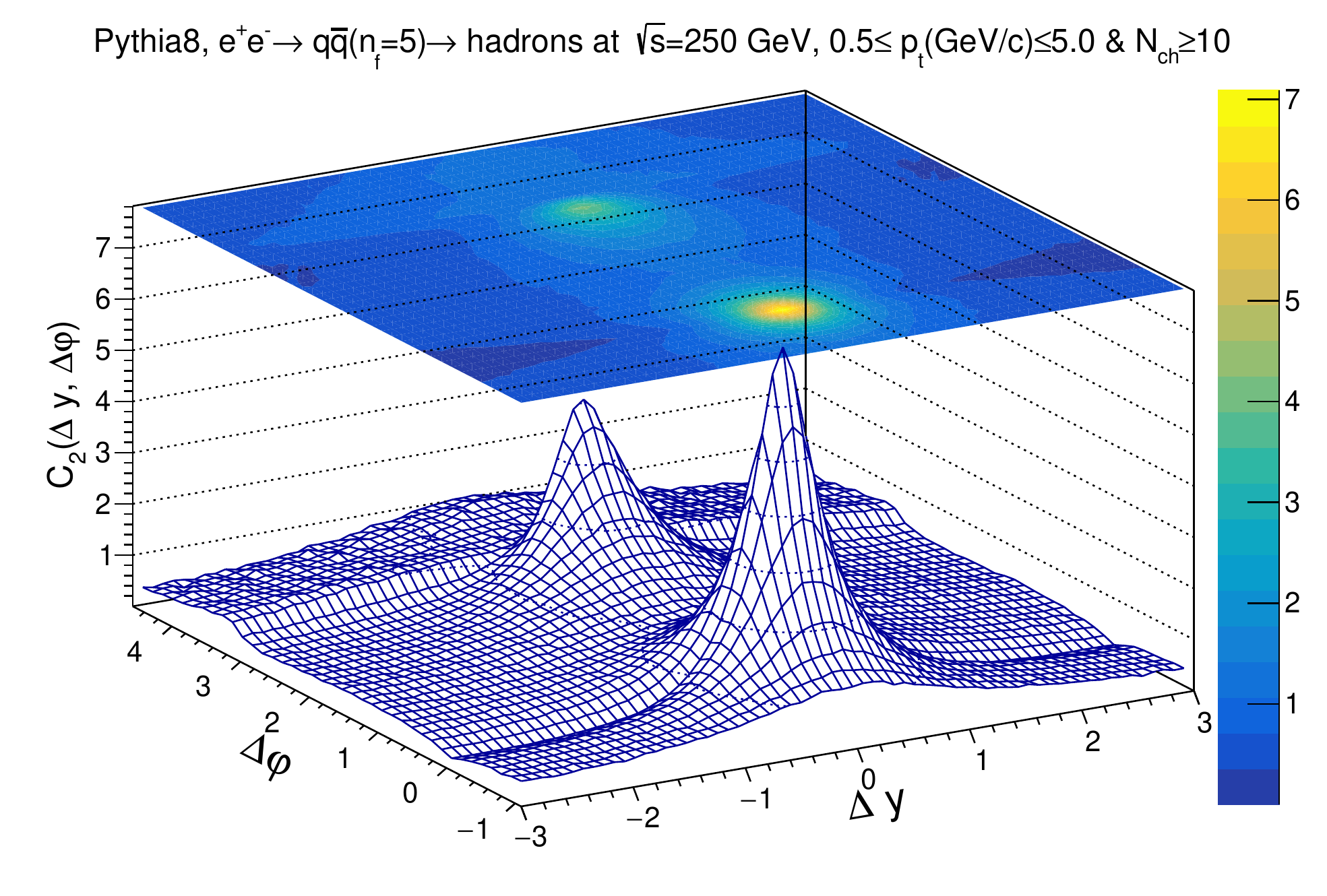, height=6.5truecm,width=8.5truecm}
\epsfig{file=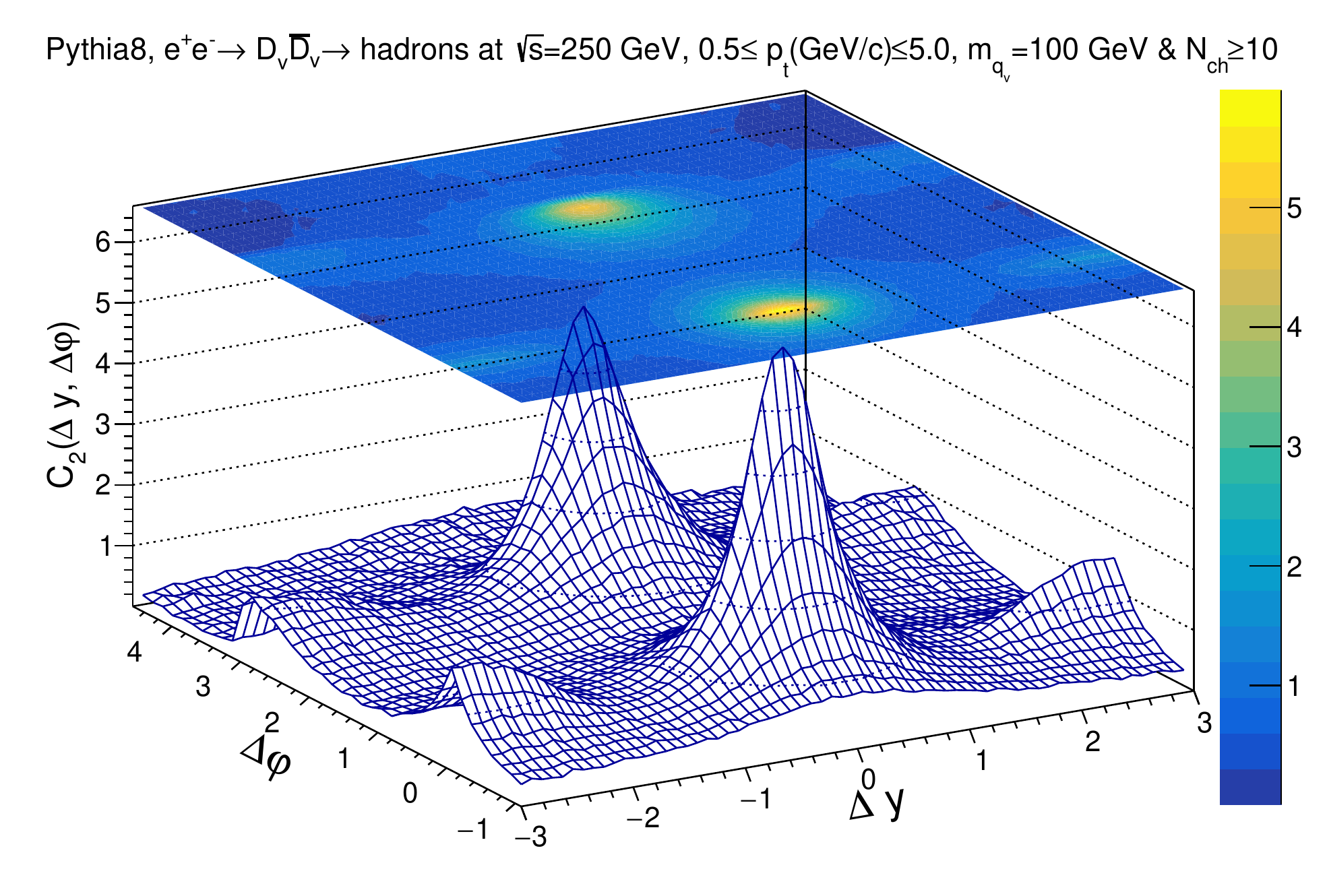, height=6.5truecm,width=8.5truecm}
\end{center}
\caption{Two-particle correlation
function, ${C}_2(\Delta y,\Delta\phi)$ (in rapidity and azimuthal angle differences), 
obtained in the SM  (left panel) and the HV scenario with 
the v-quark mass  $m_{q_v}=100$ GeV (right 
panel) in the $e^+e^-$ annihilation generated with PYTHIA8 at center-of-mass energy 
$\sqrt{s}=
250$~GeV. 
Here, $n_{\rm f}$ is the number of SM flavours, $q$ denotes all quarks except the top 
quark, $D_v$ is the HV mirror partner to the SM down quark, $N_{\rm ch}$ is the number of the 
charged 
particles, and $p_t$ is the transverse momentum of the produced particles in the sphericity 
frame.   
}
\label{fig:2-part-corr1}
\end{figure}
The Monte Carlo study here explores  
different HV scenarios for the v-partners of the SM fermions with $m_{q_v}=100$ GeV. The 
corresponding coupling constant
$\alpha_v$ was 
set to the (default) value $\alpha_v=0.1$ since correlations prove to be quite insensitive to 
higher values. 
 Correlation functions are studied as a function of 
 (pseudo)rapidity and azimuthal angle in $e^+e^-$ collisions at center-of-mass 
energy 250 GeV using the sphericity frame. 
 This choice is especially suitable in $e^+e^-$ 
colliders~\cite{OPAL:1999iqi,Grupen:1996di}
since the sphericity axis coincides with the averaged outgoing direction
of the back-to-back $q\bar{q}$ jets produced in the annihilation process.
%
%

Figure~\ref{fig:2-part-corr1} shows 3-dimensional plot of the 2-particle 
correlation function $C_2(\Delta y,\Delta \phi)$ for the SM $e^+e^-\to\gamma^*/Z^0\to q\bar{q}\to$ hadrons and the 
HV $e^+e^-\to\gamma^*/Z^*\to D_{v}\overline{D}_{v}\to$ hadrons scenarios, where the products of the decay $D_v\to d+q_v$ initiate a parton shower. The 2-particle correlation function 
represents the ratio of
the number of charged particle pairs from the same event (signal), and that 
from different events (background),
$
C_2(\Delta y,\Delta \phi)={S_2(\Delta y,\Delta \phi)}/{B_2(\Delta 
y,\Delta \phi)}
$, with the differences on rapidity and azimuthal angles of particles 1 and 2,
$\Delta y\equiv\Delta y_{12}=y_1-y_2$ and $\Delta \phi\equiv\Delta\phi_{12}=\phi_1-\phi_2$ as 
used in experiments \cite{CMS:2010ifv,PHOBOS:2008yxa,Badea:2019vey,Belle:2020mdh,ATLAS:2015hzw}. 
A nearside ridge can be expected in 
such a 3-dimensional plot of $C_2$ function,
very much like the one appearing in 
hadronic collisions. 

One can see the narrow peak for $\Delta 
\eta
\approx 0$ and $\Delta \phi \approx 0$ arising from particle pairs inside the same events, 
and the away-side ridge (at $\Delta \phi \approx \pi$) typically due to back-to back jet 
production as a consequence of energy-momentum conservation. 
 It is worth mentioning that the latter clearly exhibits stronger correlations in the HV 
sector than those 
in the SM, as visible 
from  
the plots. Note that 
the essential features of the equivalent plot, obtained 
by ALEPH~\cite{Badea:2019vey}
 and Belle~\cite{Belle:2020mdh}
experiments
are reproduced while using the thrust axis. 
\begin{figure}[!htbp]
\begin{center}
\epsfig{file=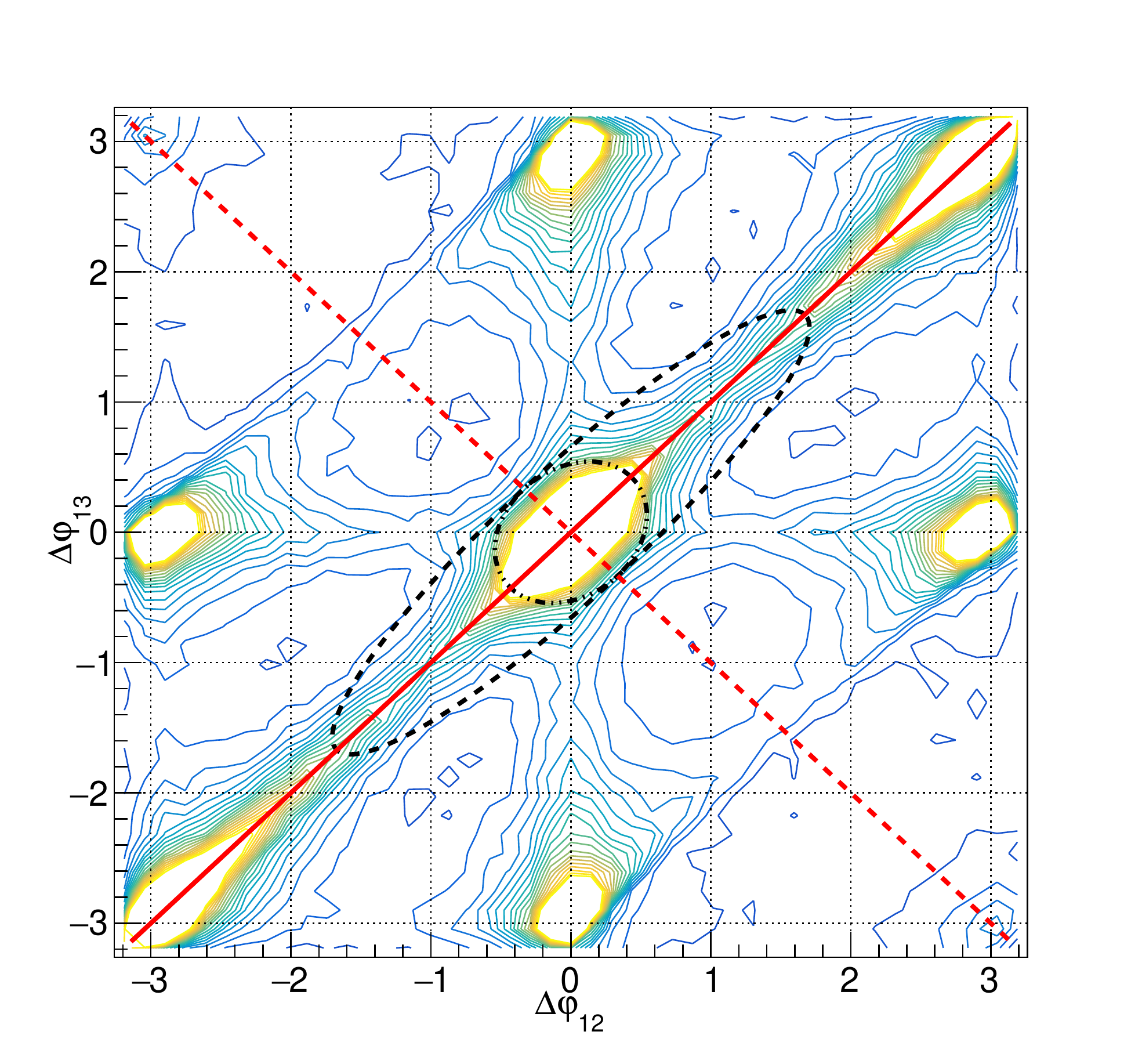, height=6.0truecm,width=6.0truecm}
\hskip -0.6cm
\epsfig{file=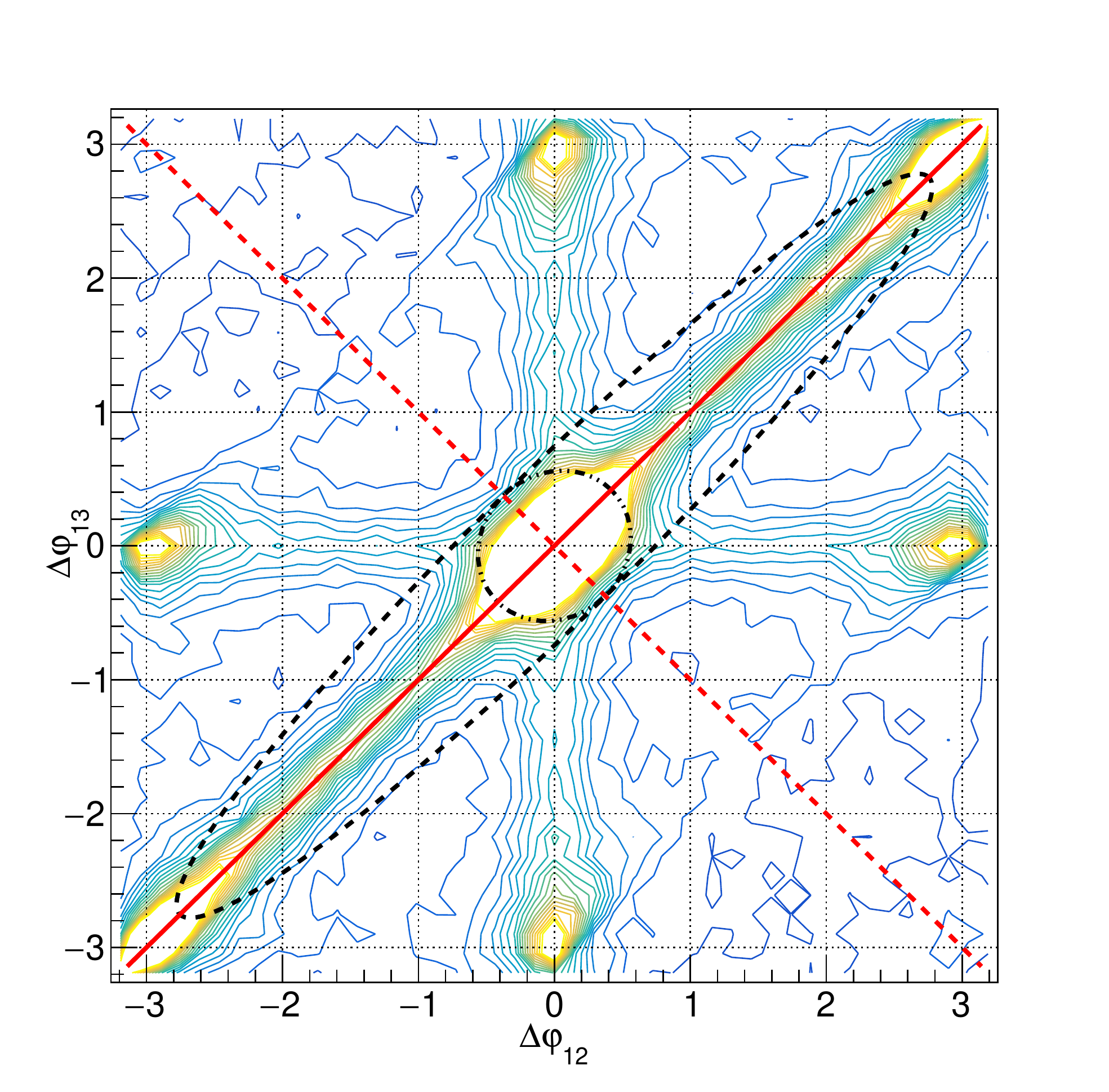, height=6.0truecm,width=6.0truecm}
\hskip -0.6cm
\epsfig{file=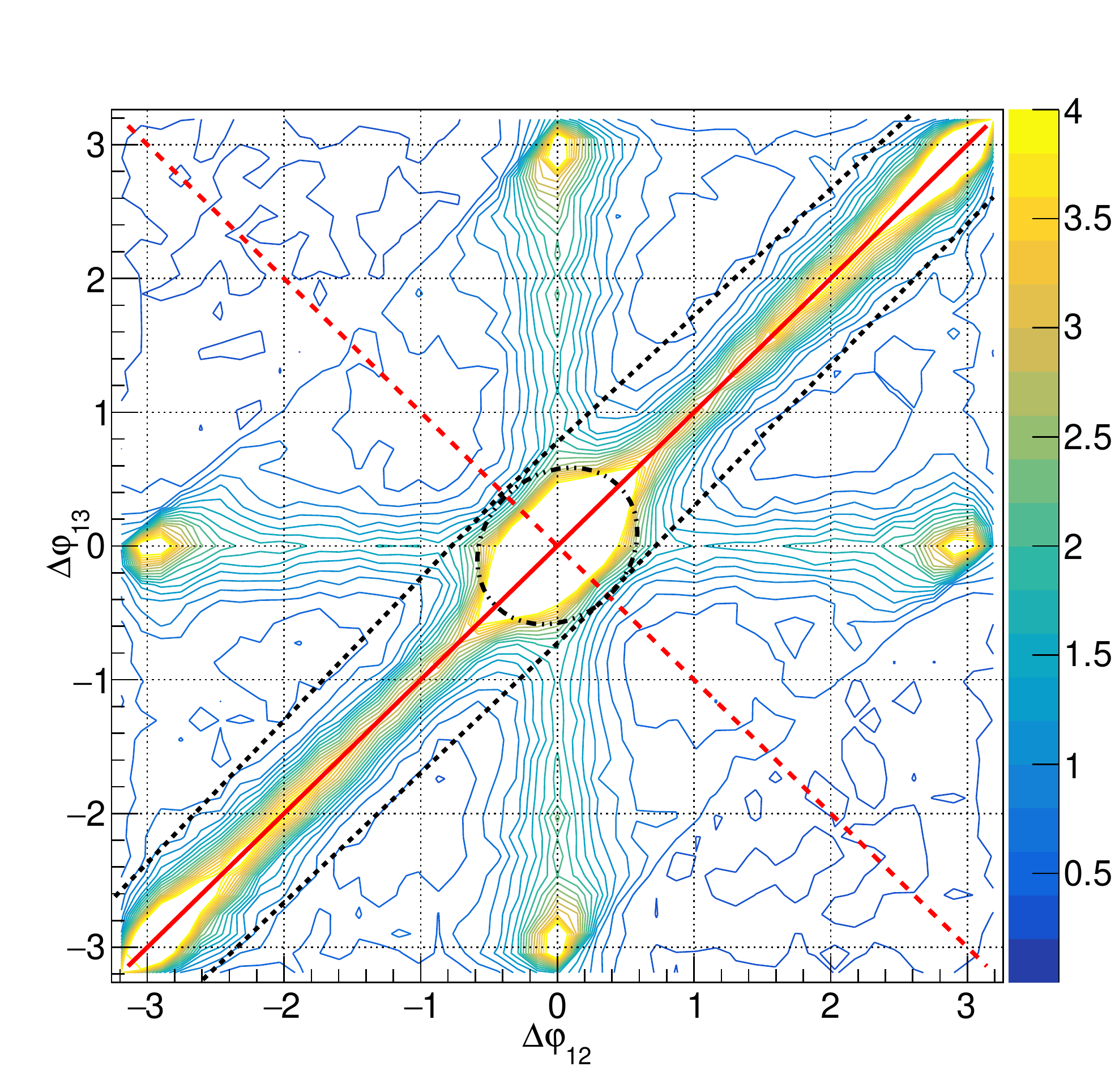, height=6.0truecm,width=6.2truecm}
\end{center}
\caption{ 
Contourplot of 3-particle azimuthal correlations 
$C^{(3)}(\Delta\phi_{12},\Delta\phi_{13})$ of charged hadrons in 
light quark (left panel), $t\bar{t}$ (middle panel) and $T_v\bar{T}_v$  
(right panel) events in $e^+e^-$ collisions generated with PYTHIA8 at $\sqrt{s}=800$ GeV. 
The solid and dotted red lines represent the on-diagonal ($\Delta\phi_{12}=\Delta\phi_{13}$) 
and off-diagonal ($\Delta\phi_{12}=-\Delta\phi_{13}$) projections, 
respectively. 
Ellipses indicate long-range correlations (major axis) and middle-range correlations (minor axis).
Long-range 
correlations increase from left to right while short-range  
and middle-range  
correlations remain as eccentricities increase. The central peak from jet correlations 
is cut off by the horizontal plane 
$C^{(3)}(\Delta\phi_{12},\Delta\phi_{13})=4$, to better illustrate the correlation structure outside that region.
}
\label{fig:3-part-corr2}
\end{figure}

Meantime, the most striking difference between both scenarios lies on the existence of long-range, nearside angular correlations (ridge~\cite{ATLAS:2015hzw}) within 
$2<|\Delta y|<4$ at $\Delta\phi\approx 0$ for the HV scenario. This
structure was checked to hold to different extends for all $F_v$ cases, while 
only the $D_v$ is plotted. 
 Such a ridge is not obtained from the SM cascade, so it 
becomes particularly  
suggestive to interpret
the ridge phenomenon as a consequence of the presence of a new stage of matter on top of
the QCD parton cascade. This is one of the main assumptions of this study,
which is examined below using 3-particle correlations.


As  is 
known \cite{highorder}, 3-particle (and higher-order) 
correlations provide extra information on the formation and evolution of 
matter in high-energy collisions. 
In what follows, 
the 
function of four variables:
$C_3(\Delta y_{12},\Delta y_{13};\Delta \phi_{12},\Delta  \phi_{13})=$ ${S_3(\Delta y_{12},\Delta y_{13};\Delta 
\phi_{12},\Delta  \phi_{13})}/$
${B_3(\Delta y_{12},\Delta y_{13};\Delta \phi_{12},\Delta  \phi_{13})}$
is used to study three-particle correlations. 
Again, the numerator represents the signal as it takes into account the number of  
particle triplets from the same 
event, while the denominator corresponds to the background, i.e. particle triplets from 
different events. 

The present study is 
restricted 
to azimuthal correlations following the findings of 
\cite{Sanchis-Lozano:2018wpz} where we concluded 
that azimuthal distributions 
are more sensitive to the presence of hidden sectors than (pseudo)rapidity distributions. 
Therefore, and for the sake of simplicity, 
the rapidity difference 
is fixed here to the interval $\Delta 
y_{12}=\Delta y_{13}=0$, that is, particles moving with similar longitudinal rapidity (along 
the sphericity axis of each event). 

Figure~\ref{fig:3-part-corr2} displays
the contour plot of  
$C_3(\Delta y_{12}=0,\Delta y_{13}=0;\Delta \phi_{12},\Delta  \phi_{13})$
for 
light quarks, 
$t\bar{t}$
and 
$T_v\overline{T}_v$ production with $\alpha_v=0.1$.
In all three cases, a rich structure can be observed
on the $(\Delta \phi_{12},\Delta  \phi_{13})$-plane:
a central high peak together with a set of $\lq\lq$satellite" peaks linked by 
a set of ridges and valleys (i.e. higher or lower values of correlations) represented as 
contour line fluxes. 
Let us 
focus 
on the central $(\Delta \phi_{12}, \Delta  \phi_{13})$-region, where the effects of the hidden 
sector mainly show up, while the other peaks can be interpreted as back-to back correlations of particle 
pairs. 
%
\begin{figure}[!htbp]
\begin{center}
\epsfig{file=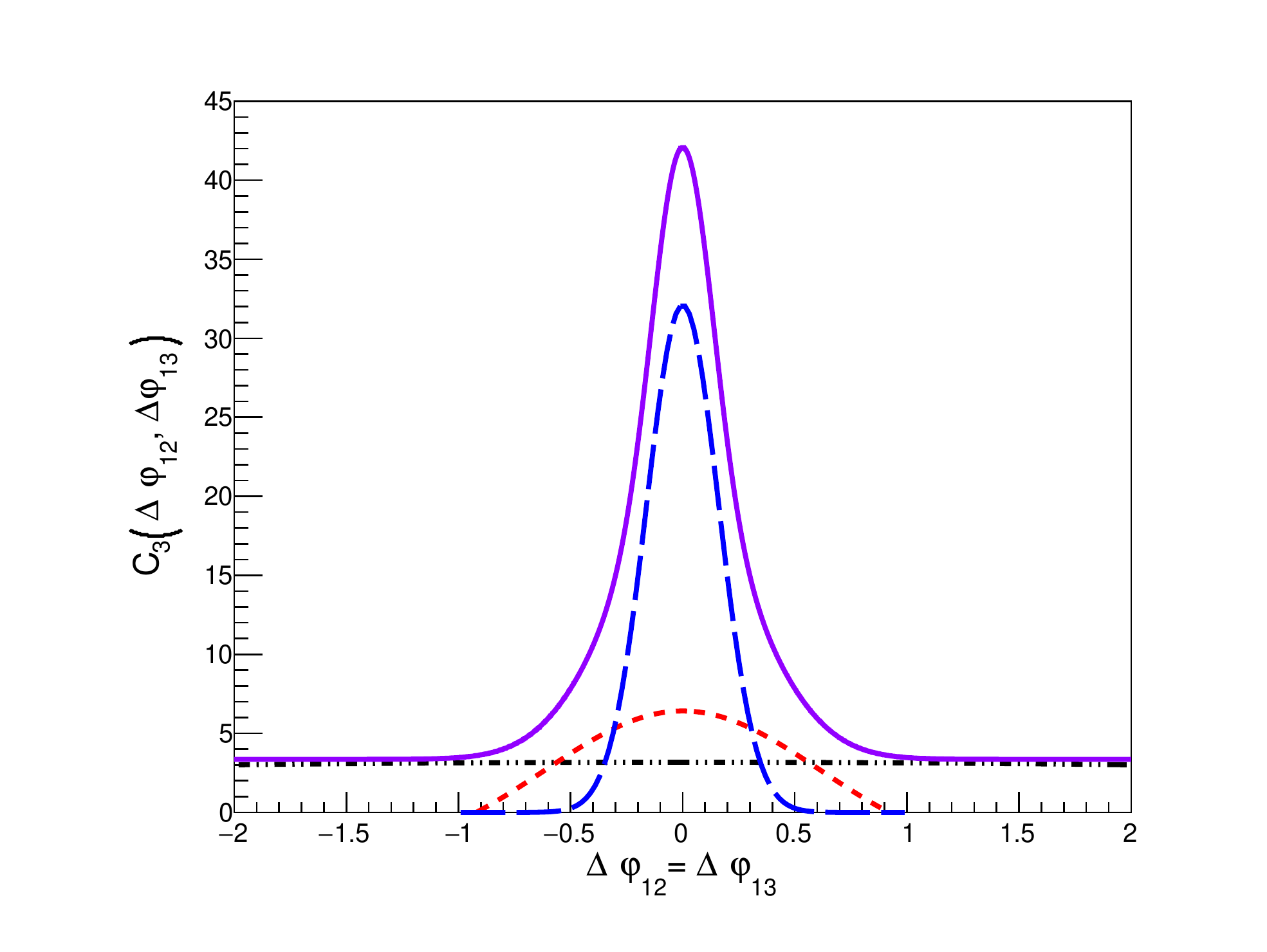, height=7.6truecm,width=8.6truecm}
\epsfig{file=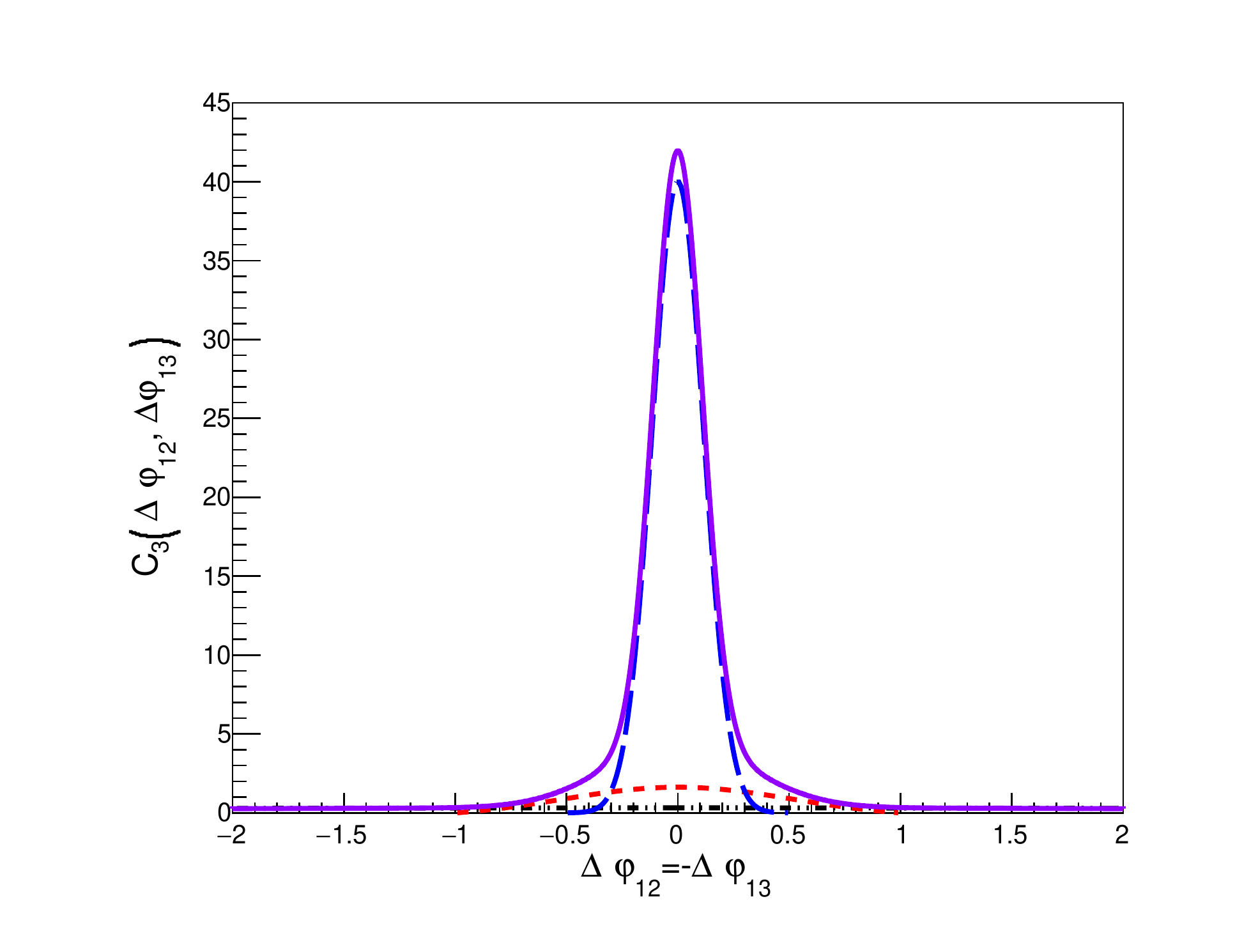, height=7.6truecm,width=8.6truecm}
\end{center}
\caption{On-diagonal (left panel) and off-diagonal (right panel) projections of the azimuthal contour-plot 
of $C_3$-function  
in the HV scenario 
obtained from Fig.\ref{fig:3-part-corr2}. 
The solid violet line represents 
a global fit 
obtained from 
 short-range correlations (long-dashed blue line), 
middle-range correlations (dashed red line) and 
long-range correlations (dotted black line).}
\label{fig:3-part-corron}
\end{figure}

Examining all three patterns shown in Fig.\ref{fig:3-part-corr2}, one can immediately 
notice 
that long-range azimuthal correlations are stretched when passing from light quarks to top 
production and furthermore to the HV sector as areas of in-middle valleys shrink 
perceptively.

To investigate in more detail such a behaviour,
one-dimensional distributions obtained from the 
on-diagonal ($\Delta \phi_{12}=\Delta \phi_{13}$) and off-diagonal ($\Delta \phi_{12}=-\Delta 
\phi_{13}$) projections of the 3-dimensional plots of Fig.\ref{fig:3-part-corr2} were plotted.
Note that long-range correlations give rise to a kind of {\it spiderweb} 
structure developing in the contour plane of the 3-particle correlation function 
\cite{Sanchis-Lozano:2017aur}. 
Fits were made 
using three Gaussian functions with 
widths, 
$\delta_s$, $\delta_m$ and $\delta_l$ corresponding, respectively, to: 
short-range correlations mainly due to resonance decays; 
middle-range correlations mainly due to the parton shower; and 
long-range correlations, attributed to the HV sector on top of the conventional cascade.
From these quantities, 
more general widths are defined 
to dissentangle the possible existence of a hidden sector, $\delta_M=\sqrt{\delta_s^2+\delta_m^2}$ and $\delta_L=\sqrt{\delta_s^2+\delta_m^2+\delta_l^2}$, such that the former takes into account the $\lq\lq$pure" QCD correlations while the latter embraces all QCD and HV correlations.

In Fig. 3, the fits for the $T_v\overline{T}_v$ case are depicted 
with three Gaussian functions for short-range, middle-range and long-range correlations along the on-diagonal and 
off-diagonal projections. Note that off-diagonal long-range correlations 
drop off faster than the on-diagonal ones, in accordance with the results of \cite{Sanchis-Lozano:2018wpz}. 
The numerical values of the widths are given in Table \ref{tab:parameters} for all three 
cases, namely, $q\bar{q}$, $t\bar{t}$
and $T_v\overline{T_v}$, obtained from the Gaussian fits to the corresponding on-diagonal  and 
off-diagonal projections 
of the $C_3$-function.  Also, 
the mean sphericity values 
are given in Table \ref{tab:parameters}
which are found to 
be the
highest (``sphere-like" events) 
for the HV events
and the 
lowest (``pencil-like" events) 
for the lightest quarks, 
while the top quark events stay in between. 
 In other words, the back-to-back hadron jets produced from 
the HV sector reach the most sphere-like shape as particles are emitted more isotropically in the event.
Then, $\Delta\phi_{12}\approx{\pi}/{2}$ or $\Delta\phi_{13}\approx{\pi}/{2}$, assuming particle 1 has been 
emitted along the sphericity axis (for a recent study see \cite{Sas:2021pup}).

\begin{table*}
\setlength{\tabcolsep}{0.5pc}
\caption{Widths (in rad) obtained from Gaussian fits of
middle-range and long-range correlations 
contributing to  $C_3(\Delta y_{12}=0,\Delta y_{13}=0;\Delta \phi_{12},\Delta  \phi_{13})$ on-diagonal and 
off-diagonal projections (Fig.\ref{fig:3-part-corron}), along with the mean sphericity values.} 

\label{tab:Tab1}

\begin{center}
\begin{tabular}{cccc}
  \hline (v)-quark species: &  light quarks  &  $t\bar{t}$ &
$T_v\overline{T}_v$ \\

\hline 
$\delta_L$(on-diagonal) & $2.2$  & $3.9$ & $6.2$   \\
\hline  
$\delta_M$(on-diagonal) & $0.55$  & $0.59$ & $0.64$   \\
\hline 
$\delta_M$(off-diagonal) & $0.44$  & $0.53$ & $0.53$  \\
\hline
mean sphericity & $0.04$  & $0.17$ & $0.47$  \\
\hline
\end{tabular}
\label{tab:parameters}
\end{center}
\end{table*}

Notice the following hierarchy involving the long-range widths shown in Table 1:
$\delta_L(HV) >\delta_L(t\overline{t})>\delta_L(q\bar{q})$, 
highlighting the systematic enhancement of long-range correlations for non-standard HV-initiated cascade compared to a standard one. This difference also matches the highest mean sphericity obtained for HV events. In its turn, $t\bar{t}$ production follows the same trend from $q\bar{q}$ to HV events. This is actually not a surprise because as said above, top-quark behaves somehow like a new step from the correlation point of view.

It is also worthwhile mentioning the observed {\it universality} showing up for 
middle-range azimuthal correlations 
by comparing the values of $\delta_M$ for all three channels, which 
are quite close. This can be understood since short-range and middle-range correlations are mainly
due to a conventional QCD shower, common to all channels regardless of
the existence of a hidden sector on top of it.

To illustrate pictorially all the above statements, 
the ellipses, whose minor/major axes numerically coincide with the 
$\delta_{M}$/$\delta_{L}$ values, 
are drawn on the contour-plot plane in
Fig.\ref{fig:3-part-corr2}.
Indeed, the similar shape and size of 
the inner ellipses is a consequence of the universality of the conventional cascade, 
while longer correlations imply larger eccentricity ellipses revealing the existence 
of a hidden sector. All these results are in overall agreement with our 
previous theoretical findings developed 
for hadronic collisions \cite{Sanchis-Lozano:2018wpz}.

To conclude, the analysis of long-range correlations in $e^+e^- \to$ 
hadrons collisions using PYTHIA8 event generator provides potentially 
observable signals on the existence of a hidden sector decaying promptly back into SM particles. The proposed signatures,
based on 2-particle and 3-particle (pseudo)rapidity/azimuthal correlations,
are complementary tools in order to further contribute to the potential discovery of HV sectors at future $e^+e^-$ colliders.\\


{\em Acknowledgments} 
This work has been partially supported by the Spanish Ministerio de Ciencia under grant PID2020-113334GB-I00/AEI/10.13039/501100011033
and by Generalitat Valenciana under grant PROMETEO/2019/113 (EXPEDITE). Redamy Pérez-Ramos is greatful to LPTHE at Sorbonne
Universit\'e, where part of this work was done.


\begin{thebibliography}{}

\bibitem{Perez-Ramos:2011jic}
R.~Perez-Ramos, V.~Mathieu and M.~A.~Sanchis-Lozano.
Three-particle correlations in QCD parton showers. 
Phys. Rev. D \textbf{84}, 034015 (2011).

\bibitem{Sanchis-Lozano:2008zjj}
M.~A.~Sanchis-Lozano.
Prospects of searching for (un)particles from Hidden Sectors using 
rapidity correlations in multiparticle production at the LHC. 
Int. J. Mod. Phys. A \textbf{24}, 4529-4572 (2009).

\bibitem{Sanchis-Lozano:2018wpz}
M.~A.~Sanchis-Lozano and E.~K.~Sarkisyan-Grinbaum.
Searching for new physics with three-particle correlations in $pp$ 
collisions at the LHC. 
Phys. Lett. B \textbf{781}, 505-509 (2018).

\bibitem{ATLAS:2015hzw}
 G. Aad {\it et al.} 
 Observation of long-range elliptic azimuthal anisotropies in
$\sqrt{s} =$~13 and 2.76
TeV $pp$ collisions with the ATLAS detector. 
Phys. Rev. Lett. \textbf{116}, 172301 (2016).


\bibitem{CMS:2010ifv}
V.~Khachatryan \textit{et al.} (CMS Collaboration).
Observation of long-range near-side angular correlations in 
proton-proton collisions at the LHC. 
JHEP \textbf{09}, 091 (2010).

\bibitem{PHOBOS:2008yxa}
B.~Alver \textit{et al.} (PHOBOS Collaboration).
System size dependence of cluster properties from two-particle angular 
correlations in Cu+Cu and Au+Au collisions at $\sqrt{s_{\rm NN}} = $~200 
GeV. 
Phys. Rev. C \textbf{81}, 024904 (2010).

\bibitem{Shuryak:2007fu}
E.~V.~Shuryak.
On the origin of the 'ridge' phenomenon induced by jets in heavy ion 
collisions, 
Phys. Rev. C \textbf{76}, 047901 (2007).

\bibitem{Dumitru:2008wn}
A.~Dumitru, F.~Gelis, L.~McLerran and R.~Venugopalan.
Glasma flux tubes and the near side ridge phenomenon at RHIC. 
Nucl. Phys. A \textbf{810}, 91-108 (2008).

\bibitem{Bozek:2012gr}
P.~Bozek and W.~Broniowski.
Correlations from hydrodynamic flow in p-Pb collisions. 
Phys. Lett. B \textbf{718}, 1557-1561 (2013).

\bibitem{Sanchis-Lozano:2020iwv}
M.~A.~Sanchis-Lozano, E.~K.~Sarkisyan-Grinbaum, J.~L.~Domenech-Garret and 
N.~Sanchis-Gual.
Cosmological analogies in the search for new physics in high-energy 
collisions. 
Phys. Rev. D \textbf{102}, 035013 (2020).

\bibitem{Carloni:2010tw}
L.~Carloni and T.~Sjostrand.
Visible Effects of Invisible Hidden Valley Radiation. 
JHEP \textbf{09}, 105 (2010).

\bibitem{ShuryakBook}
E.~V. Shuryak, {\it The QCD Vacuum, Hadrons and Superdense Matter} (World 
Scientic: Singapore, 2004).

\bibitem{Sjostrand:2014zea} 
S.~Ask, J.~R.~Christiansen, R.~Corke, N.~Desai, P.~Ilten, 
S.~Mrenna, S.~Prestel, C.~O.~Rasmussen and P.~Z.~Skands.
An introduction to PYTHIA 8.2. 
Comput. Phys. Commun. \textbf{191}, 159-177 (2015).

\bibitem{Sjostrand:2006za}
T.~Sjostrand, S.~Mrenna and P.~Z.~Skands.
PYTHIA 6.4 Physics and Manual. 
JHEP \textbf{05}, 026 (2006)

\bibitem{Sanchis-Lozano:2017aur}
M.~A.~Sanchis-Lozano and E.~Sarkisyan-Grinbaum.
Ridge effect and three-particle correlations. 
Phys. Rev. D \textbf{96}, 074012 (2017).

\bibitem{Sanchis-Lozano:2016qda}
M.~A.~Sanchis-Lozano and E.~Sarkisyan-Grinbaum.
A correlated-cluster model and the ridge phenomenon in 
hadron\textendash{}hadron collisions. 
Phys. Lett. B \textbf{766}, 170-176 (2017).

\bibitem{OPAL:1999iqi}
G.~Abbiendi \textit{et al.} (OPAL Collaboration).
Intermittency and correlations in hadronic Z0 decays. 
Eur. Phys. J. C \textbf{11}, 239-250 (1999).

\bibitem{Grupen:1996di}
C.~Grupen.
Multiparticle aspects of e$^+$e$^-$ interactions at an energy of 133 GeV 
at LEP. 
In: Proceedings on XXVI International Symposium on Multiparticle Dynamics 
(ISMD 96) (World Scientific: Singapore, 1997).
arXiv:hep-ph/9610308 [hep-ph]. 

\bibitem{Badea:2019vey}
A.~Badea, A.~Baty, P.~Chang, G.~M.~Innocenti, M.~Maggi, C.~Mcginn, 
M.~Peters, T.~A.~Sheng, J.~Thaler and Y.~J.~Lee.
Measurements of two-particle correlations in $e^+e^-$ collisions at 91 
GeV with ALEPH archived data. 
Phys. Rev. Lett. \textbf{123}, 212002 (2019).

\bibitem{Belle:2020mdh}
A.~Abdesselam \textit{et al.} (Belle Collaboration).
collisions at Belle. 
Preprint BELLE-CONF-2001, 
arXiv:2008.04187 [hep-ex]. 

\bibitem{highorder}
W. Kittel and E. A. De Wolf, E.A,
    {\it Soft Multihadron Dynamics} 
    (World Scientific: Singapore, 2005).

\bibitem{Sas:2021pup}
M.~Sas and J.~Schoppink.
Event shapes and jets in $e^+e^-$ and pp collisions. 
Nucl. Phys. A \textbf{1011}, 122195 (2021).


\end{thebibliography}


\end{document}